\documentclass[prl,twocolumn,showpacs,superscriptaddress]{revtex4}
\usepackage{graphicx}
\usepackage{amsfonts} 
\usepackage{amsmath}
\usepackage{amssymb}
\usepackage{bm}
\usepackage{slashed}
\begin{document}
\title{Kondo Signatures of a Quantum Magnetic Impurity in Topological Superconductors}
\author{Rui Wang}
\affiliation{Department of Physics and Astronomy, Shanghai Jiao Tong University, Shanghai 200240, China}
\affiliation{National Laboratory of Solid State Microstructures and Department of Physics, Nanjing University, Nanjing 210093, China}
\affiliation{Collaborative Innovation Center for Advanced Microstructures, Nanjing 210093, China}
\author {W. Su}
\affiliation{National Laboratory of Solid State Microstructures and Department of Physics, Nanjing University, Nanjing 210093, China}
\affiliation{College of Physics and Electronic Engineering, Sichuan Normal University, Chengdu 610066, China}
\author{Jian-Xin Zhu}
\affiliation{Theoretical Division, Los Alamos National Laboratory, Los Alamos, New Mexico 87545, USA}
\affiliation{Center for Integrated Nanotechnologies, Los Alamos National Laboratory, Los Alamos, New Mexico 87545, USA}
\author{C. S. Ting}
\affiliation{Department of Physics and Texas Center for Superconductivity, University of Houston, Houston, Texas 77204, USA}
\author{Hai Li}
\affiliation{Department of Physics and Texas Center for Superconductivity, University of Houston, Houston, Texas 77204, USA}
\author{Changfeng Chen}
\affiliation{Department of Physics and Astronomy, University of Nevada, Las Vegas, Nevada 89154, USA}
\author{Baigeng Wang}
\email{bgwang@nju.edu.cn}
\affiliation{National Laboratory of Solid State Microstructures and Department of Physics, Nanjing University, Nanjing 210093, China}
\affiliation{Collaborative Innovation Center for Advanced Microstructures, Nanjing 210093, China}
\author{Xiaoqun Wang}
\email{xiaoqunwang@sjtu.edu.cn}
\affiliation{Department of Physics and Astronomy, Shanghai Jiao Tong University, Shanghai 200240, China}
\affiliation{Collaborative Innovation Center for Advanced Microstructures, Nanjing 210093, China}
\affiliation{Tsung-Dao Lee Institute, Shanghai Jiao Tong University, Shanghai 200240, China}
\affiliation{Beijing Computational Science Research Center, Beijing 100084, China}

\date{\today}
\begin{abstract}
We study the Kondo physics of a quantum magnetic impurity in two-dimensional topological superconductors (TSCs), either intrinsic or induced on the surface of a bulk topological insulator, using a numerical renormalization group technique. We show that, despite sharing the $p+ip$ pairing symmetry, intrinsic and extrinsic TSCs host different physical processes that produce distinct Kondo signatures. Extrinsic TSCs harbor an unusual screening mechanism involving both electron and orbital degrees of freedom that produces rich and prominent Kondo phenomena, especially an intriguing pseudospin Kondo singlet state in the superconducting gap and a spatially anisotropic spin correlation. In sharp contrast, intrinsic TSCs support a robust impurity spin doublet ground state and an isotropic spin correlation. These findings advance fundamental knowledge of novel Kondo phenomena in TSCs and suggest experimental avenues for their detection and distinction.
\end{abstract}

\maketitle

The Kondo problem, which treats a magnetic impurity in metals \cite{JK}, is a prominent topic in materials research, and its solution by the renormalization group method invokes some of the most profound concepts and techniques in theoretical physics \cite{wilson}. Kondo phenomena offer insights into impurity scattering and screening processes and reveal characters of host materials. When a magnetic impurity is coupled to electrons in a superconductor (SC), a Yu-Shiba-Rusinov impurity state emerges in the superconducting gap \cite{lyu,hshiba}, reflecting the nature of the SC ground state \cite{Sakai,ksatori,muller,Balatsky,Matsumotoa,Matsumotob,hchen}. This idea has been extended to classical impurities in topological superconductors (TSCs) \cite{jay,Kaladzhyan,Kaladzhyanb,hhu,ttzhou,aazyuzin,Pientka}, which attract great interest \cite{msato,mleijnse} because they exhibit novel physics \cite{Kitaev,Lutchyn,Oreg,ysun,tmeng,Hor,Wray,liangf} and hold promise for topological quantum computation \cite{xlqi,Leijnse,Nayak}. Among them, two-dimensional (2D) TSCs were predicted to exist by proximity effect on the surface of a topological insulator (TI) \cite{lfuf}, which was realized in a $\mathrm{Bi}_2\mathrm{Te}_3$/$\mathrm{NbSe}_2$ heterostructure \cite{hsun}. Similar phenomena have been studied in unconventional superconductors $\mathrm{FeTe}_{0.55}\mathrm{Se}_{0.45}$ \cite{gangxu,pzhang,jxyin,hhuang} and $\mathrm{PbTaSe}_2$ \cite{gbian,syguan}. The proximate s-wave SC mediates an induced $p+ip$ paired TSC state \cite{jalicea}, and signatures of Majorana modes have been observed \cite{hsun,qlhe}. Intrinsic TSC states also have been explored in layered compound $\mathrm{Sr_2RuO_4}$ \cite{Ishida,Rice,Luke,Nelson,Mackenzie}. While these extrinsic and intrinsic TSCs share the $p+ip$ pairing, they bear fundamental differences in physical properties and underlying mechanisms \cite{jaalicear,Schnyder}.

In this Letter, we study new Kondo physics of a quantum magnetic impurity coupled to intrinsic or extrinsic 2D TSCs using a numerical renormalization group (NRG) technique. We unveil salient features of the ground state of the quantum magnetic impurity in different TSC environments. The Kondo phenomena in extrinsic TSCs are formally equivalent to those in an s-wave superconductor, but unique electronic and orbital coupling schemes drive a distinct screening mechanism that produces new Kondo features, especially a pseudospin singlet state in the superconducting gap and a spatially anisotropic spin correlation. In stark contrast, intrinsic TSCs host a spin doublet ground state and an isotropic spin correlation. These properties define new types of Kondo physics in TSCs and allow experimental distinction of TSCs driven by different pairing mechanisms.

Kondo physics is characterized by key quantities such as ground-state symmetry, impurity local density of states (LDOS), conduction electron-impurity spin correlation, and impurity susceptibility \cite{xyfeng,hewson}. We consider the Anderson impurity model in its standard form,
\begin{equation}
\hat H = \hat H_{imp} + \hat H_{hyb},
\end{equation}
\begin{equation}\label{eqssimp}
\hat H_{imp}=\sum_{\sigma}E_{f}\hat f^{\dagger}_{\sigma}\hat f_{\sigma}+U\hat n^f_{\uparrow}\hat n^f_{\downarrow},
\end{equation}
\begin{equation}\label{eqsshyb}
\hat H_{hyb}=V\sum_{\mathbf{k}\sigma}[\hat f^{\dagger}_{\sigma}\hat c_{\mathbf{k}\sigma}+\hat c^{\dagger}_{\mathbf{k}\sigma}\hat f_{\sigma}].
\end{equation}
Here $E_f$ is the local orbital energy and $U$ the Hubbard term, and we take the symmetric case, $E_f=-U/2$, which can be easily generalized to asymmetric cases \cite{Tomoki}. The hybridization term is assumed to be independent of momentum and spin.

The intrinsic TSCs with a spinful $p+ip$ pairing symmetry are described by \cite{Volovik}
\begin{equation}\label{eq0}
\begin{split}
   \hat H^{i}_0&=-\int d\mathbf{r}\hat\psi^{\dagger}(\mathbf{r})[\nabla^2/(2{m})-\mu]\hat \psi(\mathbf{r})\\
   &+\int d\mathbf{r}\frac{\Delta}{2}[\hat\psi(\mathbf{r})\sigma^y(\hat{\partial}_x+i\hat{\partial}_y)\boldsymbol{d}\cdot
   {\boldsymbol{\sigma}}\hat\psi(\mathbf{r})
   +h.c.],
\end{split}
\end{equation}
where the first part is the kinetic energy of electrons with mass $m$, and a two-component spinor annihilation operator is defined by
$\hat \psi(\mathbf{r})=[\hat c_{\uparrow}(\mathbf{r}),\hat c_{\downarrow}(\mathbf{r})]^{\mathrm{T}}$;
the second part is the pairing energy with a gap $\Delta$ and spin operator ${\boldsymbol{\sigma}}$.
The operators $\hat{\partial}_x+i\hat{\partial}_y$ ensure a $p+ip$ pairing, while the vector $\mathbf{d}$ defines an axis,
about which $\hat H^i_0$ is invariant under the transform $\hat\psi\rightarrow e^{i\theta{\mathbf{d}}\cdot{\boldsymbol{\sigma}}}\hat\psi$ in spin space.

For extrinsic TSCs, the TI surface state with a proximate $s$-wave pairing is described by
\begin{equation}\label{eq1}
\begin{split}
  \hat H^{e}_{0}&=\int d\mathbf{r}\sum_{\sigma\sigma^{\prime}}\hat c^{\dagger}_{\sigma}(\mathbf{r})[v_F(-i\boldsymbol{\sigma}\cdot\nabla)_{\sigma\sigma^{\prime}}-\mu]\hat c_{\sigma^{\prime}}(\mathbf{r})
  \\
  &+\int d\mathbf{r}\Delta [\hat c^{\dagger}_{\uparrow}(\mathbf{r})\hat c^{\dagger}_{\downarrow}(\mathbf{r})+h.c.].
\end{split}
\end{equation}
The Dirac-cone state endows an effective $p+ip$ symmetry to the pairing term after a unitary transformation.

We now present the Hamiltonian terms in the orbital angular momentum (OAM) space via $\hat c_{\mathbf{k}\sigma}=\frac{1}{\sqrt{2\pi k}}\sum_{m}e^{im\phi}\hat c_{m,\sigma}(k)$ with $m$ denoting the OAM,  $\phi$ being the angle of ${\bf k}$ respect to $x$-axis, and take $\boldsymbol{d}=\boldsymbol{e}_z$\cite{bulla,ksatori}. In this new representation,
\begin{equation}\label{eq1p}
\begin{split}
&\hat H^{i}_{0}=\int^{\infty}_0dk\{\sum_{m,\sigma}(k^2/2m-\mu)\hat c^{\dagger}_{m,\sigma}(k)\hat c_{m,\sigma}(k)\\
&~~~~+\Delta [\hat c_{0,\uparrow}(k)\hat c_{-1,\downarrow}(k)+\hat c_{0,\downarrow}(k)\hat c_{-1,\uparrow}(k)+h.c.]\}\\
&\hat H^{i}_{hyb}=\frac{\sqrt{N}V}{\sqrt{2\pi}}\int^{\infty}_0  dk\sqrt{k}\sum_{\sigma}(\hat f^{\dagger}_{\sigma}\hat c_{0,\sigma}+h.c.),
\end{split}
\end{equation}
where the intrinsic pairing is between the $m=-1$ and $0$ orbits, while the impurity is coupled only to the $m=0$ orbit. The system always stays in a spin doublet ground state (DGS) even when the Kondo effect dominates over the SC. Once Cooper pairs break up by impurity scattering in the Kondo regime, only those electrons with $m=0$ form the Kondo singlet with the {\it f}-electron, while $m=-1$ electrons are unpaired, contributing a doubly degenerate spin state coexisting with the Kondo singlet \cite{Matsumotob}.

In the extrinsic case, the Dirac cone term in Eq. \eqref{eq1} locks electronic ${\boldsymbol{\sigma}}$ and ${\bf k}$. We introduce $\gamma_{\mathbf{k},\pm}=(c_{\mathbf{k},\uparrow}\pm c_{\mathbf{k},\downarrow}e^{-i\phi})/\sqrt{2}$ to combine the spin-up and spin-down electrons. In the OAM space, using $\hat \gamma_{\mathbf{k},\alpha}=\frac{1}{\sqrt{2\pi k}}\sum_{m}e^{im\phi}\hat \gamma_{m,\alpha}(k)$, we have
\begin{equation}\label{eq8}
\begin{split}
 &\hat H^e_{0}=\int^{\infty}_0dk\{\sum_{m, \alpha}(\epsilon_{k\alpha}-\mu)\hat \gamma^{\dagger}_{m,\alpha}(k)\hat \gamma_{m,\alpha}(k)\\
&~~+\Delta[\hat \gamma^{\dagger}_{0,+}(k)\hat \gamma^{\dagger}_{-1,+}(k)+\hat \gamma^{\dagger}_{-1,-}(k)\hat \gamma^{\dagger}_{0,-}(k)]+h.c.\},\\
 &\hat H^{e}_{hyb}=\frac{\sqrt{N}V}{2\sqrt{\pi}}\int^{\infty}_0dk\sqrt{k}\{[\hat f^{\dagger}_{\uparrow}(\hat \gamma_{0,+}(k)+\hat \gamma_{0,-}(k))\\
&~~+\hat f^{\dagger}_{\downarrow}(\hat \gamma_{-1,+}(k)-\hat \gamma_{-1,-}(k))]+h.c.\},
\end{split}
\end{equation}
where  $m=-1,0$ and $\epsilon_{k \alpha}=\alpha v_Fk$. Eqs. \eqref{eq1p} and \eqref{eq8} contain both $m=-1$ and $0$ orbits as essential components in $\hat H^{e}_0$ and $\hat H^{i}_0$, reflecting the $p+ip$ pairing symmetry in TSCs, while only $m=0$ orbit is relevant for conventional s-wave SCs \cite{bulla,Sakai,ksatori}.  Moreover, while $\hat H^{i}_{hyb}$ has only $m=0$ electrons directly coupled to $\hat f_\sigma$ in Eq. \eqref{eq1p}, $\hat H^{e}_{hyb}$ evokes both $m=-1$ and $0$ electrons by the unitary transformation. This $m$-dependence of hybridization offers a sensitive probe into the TSC states (see below).

The equations in \eqref{eq8} show that the $f_\sigma$-operator is coupled to composite fermion operators $\hat{\gamma}_{0,+}(k)+\hat{\gamma}_{0,-}(k)$ and $\hat{\gamma}_{-1,+}(k)-\hat{\gamma}_{-1,-}(k)$, involving electrons from conduction and valence bands in both spin directions. We therefore introduce a new set of fermionic operators
\begin{eqnarray}
 \hat d_{\epsilon,\uparrow} &=& \frac{1}{\sqrt{2}}[\hat \gamma_{0,+}(\epsilon)\theta(\epsilon)+\hat \gamma_{0,-}(\epsilon)\theta(-\epsilon)], \\
 \hat d_{\epsilon,\downarrow} &=& \frac{1}{\sqrt{2}}[\hat \gamma_{-1,+}(\epsilon)\theta(\epsilon)-\hat \gamma_{-1,-}(\epsilon)\theta(-\epsilon)],
\end{eqnarray}
where $\theta(x)$ is the Heaviside step function and the subscript $\tau$ in $\hat d_{\epsilon,\tau}$ defines a pseudospin, which comprises electronic spin, OAM and band degrees of freedom from conduction electrons. This implies that two spin-locked bands described by Eq.\eqref{eq1} now are decoupled into two effective, independent bands, each of which is characterized by a pseudospin, but only one effective band is relevant to the Kondo process \cite{explain}. Below we use an energy representation \cite{zitko}, where the sum of $\mathbf{k}$ is converted to an integral over energy $\epsilon$ in $[-1,1]$ with the cutoff taken as the energy unit \cite{EU}, resulting in an effective Hamiltonian
\begin{equation}\label{eq13}
\begin{split}
&\hat   H^e_0=\int^{1}_{-1}d\epsilon[\sum_{\tau}g(\epsilon)\hat d^{\dagger}_{\epsilon,\tau}\hat d_{\epsilon,\tau}+\Delta(\hat d^{\dagger}_{\epsilon,\uparrow}\hat d^{\dagger}_{\epsilon,\downarrow}+h.c.)]\\
&\hat H^e_{hyb}=\int^{1}_{-1}d\epsilon h(\epsilon)\sum_{\tau}[f^{\dagger}_{\tau}\hat d_{\epsilon,\tau}+\hat d^{\dagger}_{\epsilon,\tau}\hat f_{\tau}],
\end{split}
\end{equation}
where $g(\epsilon)=\epsilon-\mu$, $h(\epsilon)=V{\rho}^{1/2}(\epsilon)/\sqrt{2}$ and ${\rho}(\epsilon)=N|\epsilon|$ $/2\pi v^2_F$ is the density of states for $\hat c_{{\bf k},\sigma}$ electrons in the TI surface. The Anderson impurity coupled to an extrinsic TSC with the $p+ip$ pairing is therefore formally equivalent to an impurity coupled to an s-wave SC described by the equations in \eqref{eq13}. There are, however, some key distinctions \cite{Sakai,ksatori}, e.g., the effective hybridization $h^2(\epsilon)\propto|\epsilon|$ linearly away from the Dirac point and the pseudospin $\tau$ accounts for both spin and orbital degrees of freedom from Eqs. \eqref{eq1}. Moreover, for $\Delta=0$, equations in \eqref{eq13} reduce to those for the Anderson impurity coupled to the TI surface state \cite{zitko}.

We employed NRG techniques \cite{wilson,jgliu} to determine low-energy properties of the Anderson impurity in TSCs described by the Hamiltonians derived above, with the effective hybridization coupling $h(\epsilon)$ properly treated for NRG calculations \cite{bulla2,zitko}. For the most interesting case of $\Delta\neq0$ and $\mu>\Delta$, the on-site potentials and hopping amplitudes in a Wilson chain adopt a matrix form such that $\hat d_{\epsilon,\tau}$-fermions are allowed to hop between two nearest sites with different $\tau$ \cite{explain}, following an established logarithmic discretization and numerical diagonalization procedure \cite{wilson,bulla}. Key NRG parameters, i.e., the number of preserved states $N$, number of z-averaging $N_z$, RG scaling parameter $\Lambda$, and length of Wilson's chain $L_{max}$, are provided in the figure captions.

\begin{figure}
\includegraphics[width=\linewidth]{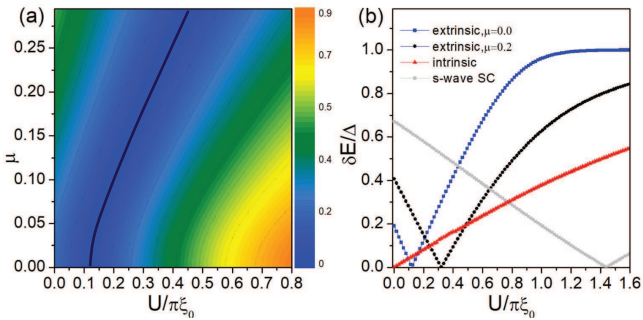}
\caption{(a) Phase diagram of $\delta E$ in $\mu$-$U$ space for extrinsic TSCs. Energy unit is set by the cutoff in Eq. (10), $\Delta=0.1$, and $U$ is reduced by $\pi\xi_0$, where $\xi_0 = \pi V^2/2$. The transition points where $\delta E=0$ are highlighted by the dark-blue line. (b) $\delta E$ versus $U$ at select $\mu$ in extrinsic TSCs ($U_c/\pi\xi_0=0.11$ and $0.32$ for $\mu=0.0$ and $0.2$, respectively), compared with results for intrinsic TSC ($U_c$=0) and s-wave SC ($U_c/\pi\xi_0=1.44$). NRG parameters used are $N$=400, $N_z$=10, $\Lambda$=2.5, $L_{max}$=25.}
\end{figure}

We first examine the lowest-excited energy relative to the ground-state energy $\delta E=E_1-E_0$ in the thermodynamic limit, which is a key quantity in probing quantum phase transitions driven by $U$ and $\mu$ \cite{Tomoki,jgliu}. The results for the impurity in extrinsic TSCs exhibit a pattern [Fig. 1(a)] showing that, for a given $\mu$, $\delta E$ initially declines with increasing $U$, reaching zero, and then rises again; each $\delta E$ versus $U$ curve [Fig. 1(b)] has a $V$ shape, and the critical value $U_c$ where $\delta E = 0$ increases with rising $\mu$. Meanwhile, $U_c$ for an impurity in a conventional s-wave SC is much higher than typical values for extrinsic TSCs, indicating that $U_c$ is suppressed by the hybridization $h(\epsilon)$ driven by the unique band structure of the proximate TI surface state in extrinsic TSCs. In contrast, intrinsic TSCs display a monotonically increasing $\delta E$ with $U_c = 0$, indicating an absence of any quantum phase transition in the $\mu$-$U$ space.

To elucidate the behavior of $\delta E$ in the context of new Kondo physics, we assess impurity moment $M_{imp}=\lim_{T\rightarrow 0} \sqrt{T\chi_{imp}(T)}$, where $\chi_{imp}(T)$ is impurity susceptibility and $T$ absolute temperature. Calculated $M_{imp}$ in the $\mu$-$U$ space is shown in Fig. 2(a). For a fixed $\mu$, the system goes from $M_{imp}=0$ at $U<U_c$ to $M_{imp}=1/2$ at $U>U_c$, corresponding to a phase transition from a spin singlet to doublet ground state. These general features of $M_{imp}$ is similar to those in a conventional s-wave SC \cite{Tomoki}; in contrast, the impurity in intrinsic TSCs stays in the DGS regime at all $\mu$ without a phase transition.

\begin{figure}
\includegraphics[width=\linewidth]{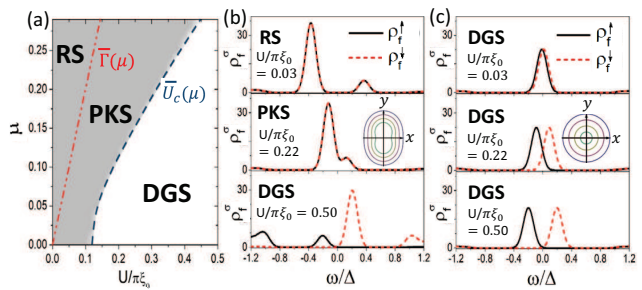}
\caption{(a) Phase diagram of $M_{imp}$ in $\mu$-$U$ space for extrinsic TSCs (same energy units as in Fig. 1 and $\Delta=0.1$). $U_c(\mu)$ sets the phase boundary between $M_{imp}=0$ and $M_{imp}=1/2$, and $\Gamma(\mu)$ is a crossover boundary between the RS and PKS regimes. Both quantities are reduced by $\pi \xi_0$, as indicated by a bar over each of them. Representative impurity LDOS at $\mu$=0.2 for (b) the RS, PKS and DGS ($S_f^z=1/2$) regimes of extrinsic TSCs and (c) intrinsic TSCs. $\omega=0$ is at the chemical potential $\mu$ of the normal state. Spectral features are broadened in a log-Gaussian scheme with a width factor $b=0.01$ \cite{bulla3}. Contour plots of spin correlation function $C^x(\mathbf{r})$ in the PKS regime for extrinsic TSCs and in the corresponding DGS regime for intrinsic TSCs are shown as insets. Contour values are -0.10 to -0.07 with a step of 0.01 outward. NRG parameters used are $N$=600, $N_z$=10, $\Lambda$=2.5, $L_{max}$=25.}
\end{figure}

At $U>U_c$, Cooper pairs formed by $\hat{d}_{\epsilon\tau}$-fermions are robust against impurity scattering, leading to a degenerate $\hat f_{\sigma}$-electron doublet state, placing the system in the DGS regime in Fig. 2(a). For $U \leq U_c$, $M_{imp}=0$, but the phase space is further divided into two areas separated by a crossover governed by a function $\Gamma(\mu)=\pi V^2\rho(\mu)$ with $\rho(\mu)$ being the DOS of the normal state at $\mu$. At $0<U\leq\Gamma$, charge fluctuations of $f$-electrons allow for resonance scattering (RS) between the impurity and conduction electrons \cite{Tomoki}; charge fluctuations are greatly suppressed when $U\gg \Gamma$ \cite{explain}, resulting in the formation of a pseudospin Kondo singlet (PKS) state.

In the PKS regime, Cooper pairs are broken by impurity scattering, and pseudospins $\tau$ of $\hat{d}_{\epsilon\tau}$-fermions released from Cooper pairs form the PKS with the impurity spin. Both the $m=-1,0$ orbits and spin of conduction electrons $\hat c_{m,\sigma}$ are involved in screening the impurity spin. This unusual process produces a spatially anisotropic correlation between the impurity and conduction electron spins [Fig. 2(b)], given by $C^k({\bf r})=\langle \hat S^{k}_c(\mathbf{r})\hat S^k_f(0)\rangle$, with $k=x,y,z$ and $\hat{\mathbf{S}}_c(\mathbf{r})=\hat c^{\dagger}(\mathbf{r})\boldsymbol{\sigma}\hat c(\mathbf{r})/2$ and $\hat {\mathbf{S}}_f(0)=\hat f^{\dagger}\boldsymbol{\sigma}\hat f/2$ \cite{explain}. This feature reflects the spin-momentum locking of the conduction electrons in TI induced extrinsic TSCs, distinguishing the PKS from the conventional Kondo singlet (KS) in normal metals.

We also evaluated impurity LDOS $\rho_{f}^{\sigma}(\omega)=-{\rm Im} G^{\sigma}_f(\omega)/\pi$ with $G^{\sigma}_f(\omega)$ being the Green's function of the $f$-electron with spin $\sigma$. We adopted established NRG schemes for spectral densities \cite{jbauer,tffang1}, employing the standard log-Gaussian broadening scheme with a width factor $b=0.01$ \cite{bulla3} on the delta-function-like in-gap spectral features to show clearly the spin degeneracy and states both inside and outside the gap. For the Anderson impurity coupled to a standard s-wave SC, the LDOS in-gap peaks are usually located near $\omega=\pm \Delta$ \cite{Tomoki} in the RS regime. Here, for the extrinsic TSC case, $\rho^{\sigma}_{f}(\omega)$ is spin-independent and has two well-separated peaks in the low-$U$ RS regime. Owing to the effective hybridization $h(\epsilon)$, these peaks move deeper inside the gap toward $\omega=0$ as seen in Fig. 2(b). At increasing $U$ below $U_c$, the system undergoes a crossover into the PKS regime, where the two peaks evolve continuously and move closer to $\omega=0$. For $U>U_c$, the ground state becomes two-fold degenerate with $S^z_f=\pm1/2$ so that Yu-Shiba-Rusinov type in-gap peaks emerge, driven by the scattering between the local moment and Cooper pairs. Different from the RS and PKS regimes, $\rho^{\uparrow}_{f}(\omega)$ and $\rho^{\downarrow}_{f}(\omega)$ have different profiles in the DGS regime as shown for the $S_f^z=1/2$ state in the bottom panel of Fig. 2(b). Meanwhile, the impurity in intrinsic TSCs remains in DGS and its LDOS is less sensitive to parameter changes [Fig. 2(c)]. These distinct LDOS behaviors should be detectable by spin-resolved scanning tunneling microscopy measurements \cite{Cornils}, thereby distinguishing extrinsic and intrinsic TSCs.

\begin{figure}
\includegraphics[width=\linewidth]{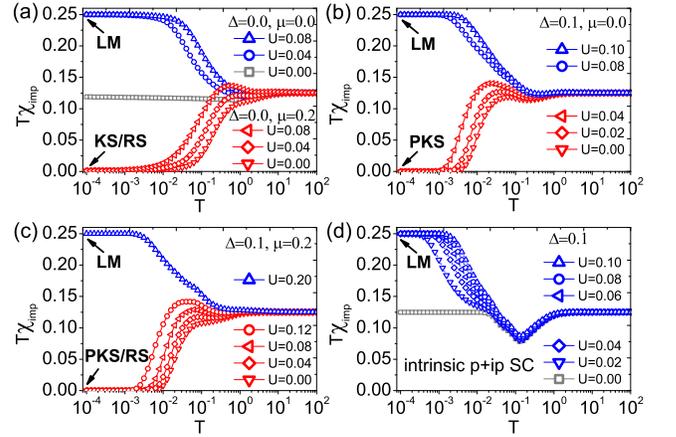}
\caption{The scaling behavior of $T\chi_{imp}(T)$ for the impurity in (a-c) extrinsic and (d) intrinsic TSCs. (a) At $\Delta=0$, the RG flow goes to a LM fixed point with $M_{imp}=1/2$ for $\mu=0$ and, when $\mu\neq0$, to KS and RS fixed points with $M_{imp}=0$ for $U\neq 0$ and $U=0$, respectively. (b) At $\Delta \neq 0$ and $\mu = 0$, the RG flow goes to $M_{imp}=1/2$ (LM) and $0$ (PKS) fixed points for $U>U_c$ and $U<U_c$, respectively. (c) At $\Delta \neq 0$ and $\mu \neq 0$, an additional RS fixed point with $M_{imp}=0$ appears. (d) RG flows go to the LM fixed point for $U\neq 0$ at all $\mu$. NRG parameters used are $N$=600, $N_z$=10, $\Lambda$=1.8, $L_{max}$=15.}
\end{figure}

We now examine the scaling behavior of $T\chi_{imp}(T)$. For extrinsic TSCs, when $\Delta=0$ the impurity is coupled to a TI surface state and not screened by conduction electrons for any $U$ at $\mu=0$ since the Fermi energy is at the Dirac point, whereas it is screened and forms a KS for any $U\neq0$ as long as $\mu \neq 0$ \cite{zitko,xyfeng}. NRG results for $\Delta=0$ in Fig. 3(a) indeed show RG flow going to a local-moment (LM) (or DGS) fixed point with $M_{imp}=1/2$ for $U\neq0$ at $\mu=0$, and to a KS fixed point with $M_{imp}=0$ for all $U$ at $\mu\neq0$. When $\Delta\neq0$, the RG flow goes to fixed points with $M_{imp}=0$ and $1/2$ for $U<U_c$ and $U>U_c$, respectively, even at $\mu=0$ as shown in Fig. 3(b) \cite{footnote}. In this case, the impurity is screened by the $\hat{d}_{\epsilon\tau}$-fermions released from the Cooper pairs for sufficiently low $T$ and $U<U_c$. When $\mu\neq0$, in addition to the fixed points at $\mu=0$, an RS fixed point with $M_{imp}=0$ appears at $U\leq\Gamma$ as shown in Fig. 3(c). Although the PKS and RS fixed point can hardly be distinguished by $T\chi_{imp}(T)$, they manifest themselves by different locations of in-gap LDOS peaks. It should be noted that NRG calculations of $T\chi_{imp}(T)$ are sensitive to computational procedures \cite{tffang2} and NRG parameters \cite{ksatori,jbauer}, especially at $T\sim \Delta$. We have performed extensive calculations and analysis to choose suitable parameters \cite{explain}. For comparison, we also show $T\chi_{imp}(T)$ for the impurity in intrinsic TSCs in Fig. 3(d), where only the LM fixed point with $M_{imp}=1/2$ exits for all $U \neq 0$. These results highlight fundamentally different Kondo physics in intrinsic and extrinsic TSCs.

In summary, we have uncovered new Kondo phenomena associated with a quantum magnetic impurity in extrinsic and intrinsic 2D TSCs with a $p+ip$ pairing symmetry. For extrinsic TSCs, a spin-momentum locking in the surface TI state generates an effective coupling between the impurity and the spin and $m=-1, 0$ orbits of conduction electrons, resulting in a rich phase diagram characterized by RS, PKS and DGS regimes. In the PKS regime, both spin and orbital degrees of freedom participate in the screening process, producing an unusual Kondo state with a spatially anisotropic spin correlation. It should be noted that this type of Kondo state will not appear in a proximity-induced SC like graphene on an s-wave SC substrate due to a lack of the spin-momentum locking. For intrinsic TSCs, the impurity stays  in a robust DGS regime with an isotropic spin correlation. These two cases are further distinguished by different evolutions of LDOS with changing $U$ and $\mu$, which reveal intriguing underlying physical processes and open avenues for experimental verification.

Recently, spin-polarized Yu-Shiba-Rusinov states were observed by spin-resolved spectroscopy measurements in a conventional s-wave superconductor \cite{Cornils}. Such experimental probes should be applicable for measuring the impurity LDOS and spin susceptibility in $\mathrm{Bi}_2\mathrm{Te}_3/\mathrm{NbSe}_2$ heterostructure \cite{hsun,qlhe}, where 2D extrinsic $p+ip$ SC state is present in the surfaces of the $\mathrm{Bi}_2\mathrm{Te}_3$ TI thin film of several quintuple layers via the proximity effect with the $\mathrm{NbSe}_2$ SC substrate. In addition, STM/STS measurements may also be properly devised to detect tunneling current of an Anderson impurity. Furthermore, $\mathrm{FeTe}_{0.55}\mathrm{Se}_{0.45}$ \cite{gangxu,pzhang,jxyin,hhuang} or $\mathrm{PbTaSe}_2$ \cite{gbian,syguan} may provide a natural platform to detect the novel Kondo effects. Meanwhile, the present studies can be generalized to a wide range of additional models coupled to a quantum magnetic impurity, such as 1D TSCs \cite{Kitaev,jli,lfukane,Rokhinson,Lutchyn,Oreg}, the quantum anomalous Hall effect with induced superconductivity \cite{xllqi,czzchang,checkelsky}, and 3D superconducting TIs and semimetals \cite{Kriener,Sasaki,mxwang,Cho}. Our work reported here is therefore expected to have major implications for further exploration of novel quantum magnetic impurity effects in a variety of exotic electronic environments involving either effective or equivalent spin-orbital couplings.

We are grateful to Jinfeng Jia, Jinguo Liu, Shuheng Pan, Qianghua Wang, Tao Xiang and Lu Yu for fruitful discussions.
This work was supported by the National Program on Key Research Project (No. 2016YFA 0300501), the National Science Foundation of China (No. 60825402, 11574217, 11204035 and 11574200), Texas Center for Superconductivity at the University of Houston, the Robert A. Welch Foundation under No. E-1146 and U.S. DOE/BES under No. LANLE3B7.

R. W. and W. S. contributed equally to this work.


\begin{thebibliography}{99}
\bibitem{EU} This energy cutoff covers the energy range of the Dirac cone state, which usually equals the bulk band gap of the 3D
topological insulator, and its value is material dependent.
\bibitem{footnote} The results for $\mu\neq0$  is examined being qualitatively the same as for $\mu=0$ when $\mu<\Delta$.
\end{thebibliography}
\end{document}